\documentclass{elsarticle}
\makeatletter
\def\ps@pprintTitle{%
 \let\@oddhead\@empty
 \let\@evenhead\@empty
 \def\@oddfoot{\centerline{\thepage}}%
 \let\@evenfoot\@oddfoot}
\makeatother

\usepackage[utf8]{inputenc}
\usepackage{amsmath, amssymb, amsbsy, amsthm, epigraph}
\usepackage{graphicx, xcolor, url, tikz}
\usepackage{scrtime, currfile}
\usetikzlibrary{fit}

\renewcommand{\phi}{\varphi}
\renewcommand{\theta}{\vartheta}

\newcommand{\calR}{\mathcal{R}}

\newcommand{\lb}{\left(}
\newcommand{\lsb}{\left[}

\newcommand{\rb}{\right)}
\newcommand{\rsb}{\right]}

\newcommand{\lpm}{\begin{pmatrix}}
\newcommand{\rpm}{\end{pmatrix}}

\begin{document}
\title{A two--strain SARS--COV--2 model for Germany - Evidence from a Linearization}
\author[1]{Thomas Götz\corref{cor1}}
\ead{goetz@uni-koblenz.de}
\address[1]{Mathematical Institute, University Koblenz, 56070 Koblenz, Germany}
\author[2]{Wolfgang Bock}
\ead{bock@mathematik.uni-kl.de}
\address[2]{Department of Mathematics, TU Kaiserslautern, 67663 Kaiserslautern, Germany}
\author[1]{Robert Rockenfeller}
\ead{rrockenfeller@uni-koblenz.de}
\author[1]{Moritz Schäfer}
\ead{moritzschaefer@uni-koblenz.de}

\cortext[cor1]{Corresponding author.}

\begin{abstract}
Currently, due to the COVID--19 pandemic the public life in most European countries stopped almost completely due to measures against the spread of the virus. Efforts to limit the number of new infections are threatened by the advent of new variants of the SARS--COV--2 virus, most prominent the B.1.1.7 strain with higher infectivity. In this article we consider a basic two--strain SIR model to explain the spread of those variants in Germany on small time scales. For a linearized version of the model we calculate relevant variables like the time of minimal infections or the dynamics of the share of variants analytically. These analytical approximations and numerical simulations are in a good agreement to data reported by the Robert--Koch--Institute (RKI) in Germany.
\end{abstract}


\maketitle


\section{Introduction}
The current COVID--19 pandemic is striking across the world and has put Europe at the dawn of its third wave. In Germany due to the rising numbers at the end of the year 2020, the non-pharmaceutical intervention (NPI) measures have been strengthened, leading to a severe lockdown with closing of the main parts of the daily life. With the reestablishment of the NPIs the reports of new strains of the SARS--COV--2 virus throughout Europe were rising \cite{Pachetti}. Especially a variant called B.1.1.7 that was first reported in Great Britain \cite{VOLZ, Tang}, showed an increased infectivity\cite{Korber} with a higher attack rate especially in the younger age groups. 
In the last days (February 21, 2021) the incidences are stagnating or slowly increasing, although Germany has not eased the lockdown. One explanation among experts and media is the rising of incidences with the new mutations. 

In various countries B.1.1.7. is rapidly spreading, see~\cite{B117data} for an overview. 
\begin{figure}
\centerline{\includegraphics[width=.8\textwidth]{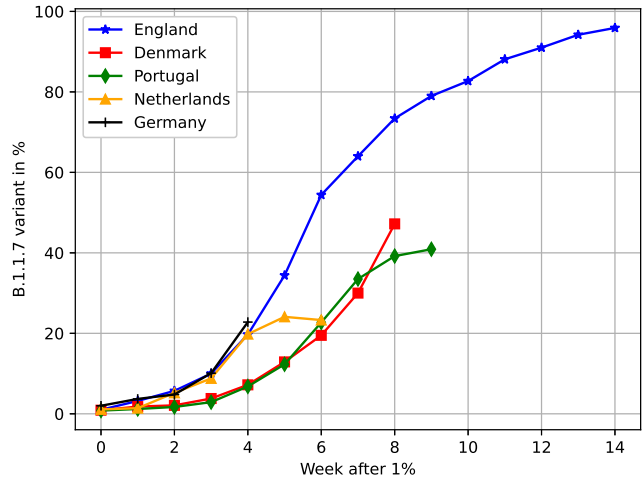}}
\caption{\label{F:countries} Share of the B.1.1.7 variant of the SARS--COV--2 virus in five European countries. Week zero corresponds to the week when the share is approximately $1\%$.}
\end{figure}
In Figure~\ref{F:countries} we show the share of this new strain with respect to analyzed SARS--COV--2–positive tests in a given week in five European countries. Week zero is defined as the week, when this share was approximately $1\%$. All five countries follow a general logistic trend. The curves for England (blue), Netherlands (orange) and Germany (black) are rather similar, where as Denmark (red) and Portugal (green) also behave similar but slower than the first three. 
Within this paper we will formulate a SIR--based model that predicts these curves and explains well the observed data in Germany on a small time horizon.

\section{Mathematical Model}

We consider an SIR--model for the spread of two strains of the SARS--COV--2 virus within a constant and serologically naive population. The two competing strains $1$ and $2$ are assumed to have  different transmission rates $\beta_2,\beta_1>0$ but the same recovery rate $\gamma>0$. Assuming no secondary infections, the model is based on the four compartments: susceptible $S$, infected $I_1$ and $I_2$, indicating strain $1$ or $2$,respectively, and removed $R$. Neglecting demographic effects like birth and death, we get
\begin{subequations}
\begin{align}
	S' &= - \lb \beta_1 I_1 + \beta_2 I_2 \rb \frac{S}{N}\;, \\
	I_1' &= \frac{\beta_1 I_1}{N} S - \gamma I_1\;, \\
	I_2' &= \frac{\beta_2 I_2}{N} S - \gamma I_2\;,\\
	R' &= \gamma (I_1+I_2)\;.
	\label{E:SIR}
\end{align} 
\end{subequations}
For the following analysis and simulations we assume a situation that models the competition between the original SARS--COV--2 virus and mutated variants like B.1.1.7 that is currently observed in many countries throughout Europe. The second (mutated) strain has a higher infection rate, i.e.~$\beta_2>\beta_1$. However, at the initial time, the original strain $1$ is still dominant in the population i.e.~$I_1(0)>I_2(0)$. Current non--pharmaceutical interventions are strict enough to suppress the original strain, i.e.~to force its reproduction number below the epidemic threshold~$\calR_1:=\frac{\beta_1}{\gamma}<1$. However the mutated strain $2$, due to its higher infectivity, might reach a reproduction number $\calR_2>1$ and hence drives the epidemic.

Typical questions that might arise in this setting:
\begin{itemize}
    \item When will strain 2 dominate the dynamics? After what time $T^\ast$ do we observe $I_2(T^\ast) > I_1(T^\ast)$?
    \item How does the total number of infected $I=I_1+I_2$ evolve in time? At what time $\check{T}$ do we observe a local minimum of the infections?
\end{itemize}

In our model, we neglect the effect of possible vaccinations, that might have different efficiency with respect to the two strains.

\section{Analysis}
Let $N$ denote the constant total population. We rescale the populations  $s=S/N$ , $x=I_1/N$, $y=I_2/N$ and $r=R/N$ and introduce a non--dimensional time $\gamma t$. Then we get
\begin{subequations}
\label{E:SIR_nondim}
\begin{align}
	s' &= - \lb \calR_1 x + \calR_2 y \rb s \\
	x' &= \lb\calR_1 s - 1 \rb x\\
	y' &= \lb \calR_2 s -1 \rb y \\
	r' &= x+y
\end{align}
\end{subequations}
where $\calR_i=\beta_i/\gamma$. Setting $y=zx$, where $z$ denotes the ratio between infected with strains $1$ and $2$, we get
\begin{subequations}
\begin{align}
    z' &= \lb \calR_2-\calR_1\rb s z
\intertext{with the solution}
    z(t) &=  z_0 \exp \lsb (\calR_2-\calR_1)\int_0^t s(t)\,dt\rsb\;.
\end{align}
\end{subequations}
In case of $\calR_2>\calR_1$, the ratio between the two strains is going to tend towards strain $2$, i.e.~$z>1$.

\subsection{Linearized Setting}

In case of dominating susceptibles, i.e.~$s\approx 1$ the ODEs~\eqref{E:SIR_nondim} linearize
\begin{subequations}
\label{E:SIR_lin}
\begin{align}
    x' &= \lb\calR_1 - 1 \rb x\\
	y' &= \lb \calR_2 -1 \rb y\;,
\end{align}
\end{subequations} 
and we are able to solve them explicitly for the infected compartments $x,y$. For both compartments we observe an exponential behavior; however since $\calR_1<1$ the compartment $x$ is dying out and compartment $y$ is exponentially growing due to $\calR_2>1$. The ratio $z$ of the two strains exhibits an exponential increase
\begin{equation}
	z(t) = z_0 e^{(\calR_2-\calR_1)t}\;.
\end{equation}
In this setting we can easily answer the initial questions posed in section 2: 
\begin{enumerate}
    \item Strain $2$ will ''overtake'' strain $1$ at time $T^\ast$, i.e. $z(T^\ast)=1$. In the linearized model~\eqref{E:SIR_lin} this time $T^\ast$ is given by
    \begin{equation}
        \label{E:T_star_lin}
	    T^\ast = - \frac{\ln z_0}{\calR_2-\calR_1}>0\;
    \end{equation}
    since $z_0<1$ and $\calR_2>\calR_1$. 
    \item The total infected attain a local minimum at time $\check{T}$ when~$(x+y)'(\check{T})=0$. In the linearized model~\eqref{E:SIR_lin} it holds that 
    \begin{equation*}
        x'+y' = (\calR_1-1)x+(\calR_2-1)zx
    \end{equation*}
    and hence $z(\check{T}) =z_0 e^{(\calR_2-\calR_1)\check{T}} = \frac{1-\calR_1}{\calR_2-1}>0$. So we arrive at
    \begin{subequations}
    \label{E:Tz_min_lin}
    \begin{align}
        \check{T} &= \frac{1}{\calR_2-\calR_1}\ln \frac{1-\calR_1}{z_0(\calR_2-1)} 
            = T^\ast\cdot  \ln \frac{1-\calR_1}{\calR_2-1}\\
        \check{z} = z(\check{T}) &= \frac{1-\calR_1}{\calR_2-1}\;.
    \end{align}
    The minimal number of infected is given by
    \begin{align}
        (x+y)_{\min} &= x_0 e^{(\calR_1-1)\check{T}}\lb 1+\check{z} \rb \notag \\
            &= x_0 \lsb \frac{1-\calR_1}{z_0(\calR_2-1)}\rsb^{(\calR_1-1)/(\calR_2-\calR_1)}\cdot\frac{\calR_2-\calR_1}{\calR_2-1}
    \end{align}
    \end{subequations}
\end{enumerate}

The relative share $p=y/(x+y)=z/(1+z)$ of the second strain $y$ with respect to the total infected $x+y$ satisfies in the linearized setting the following logistic relation
\begin{equation}
    p = \frac{z_0 e^{(\calR_2-\calR_1)t}}{1+z_0e^{(\calR_2-\calR_1)t}}
        = \frac{z_0}{z_0+e^{-(\calR_2-\calR_1)t}} \in [0,1]\;.
\end{equation}

In the linear, single strain model $x'=(\calR_1-1)x$ the reproduction number satisfies the relation
\begin{equation*}
    \calR_1 = 1+ \frac{d}{dt} \ln x\;.
\end{equation*}
Hence we may define analogously the current reproduction number $\calR(t)$ for the total infected as
\begin{align}
    \calR(t) &:= 1+ \frac{d}{dt} \ln (x+y) \;. \label{E:R_of_T}
\intertext{Using the solution of the linear model $x+y=x_0 e^{(\calR_1-1)t}+x_0 z_0 e^{(\calR_2-1)t}$ we obtain the convex combination}
    \calR(t) &:= (1-p)\calR_1 + p\calR_2 =
    \calR_1 + \frac{\calR_2-\calR_1}{1+\frac{1}{z_0} e^{-(\calR_2-\calR_1)t}} \;,  \label{E:R_of_T_lin}
\end{align}
i.e.~a logistic behavior switching between $\calR_1$ for $t\to -\infty$ and $\calR_2$ for $t\gg 1$. At time $\check{T}$, when the total number of infected attains its minimum, the current reproduction number crosses the stability threshold, i.e.~$\calR(\check{T})=1$. For the non--linear model, the overall behavior of the reproduction number is similar, despite the saturation effect due to the decreasing pool of susceptibles.

\section{Simulations}
For our simulations, we assume the following data roughly resembling the situation in Germany by mid of January to mid of February:
\begin{enumerate}
\item The total population equals to $N=83$ millions including $3$ millions of recovered or vaccinated and $x_0\simeq 78.000$ infected with variant 1 and $y_0=z_0\cdot x_0$ infected with strain 2. 
\item The recovery period is assumed to be $1/\gamma=5$ days.
\item Strain $1$ has a reproduction number of $\calR_1=0.85$, i.e. the current lockdown measures are strict enough to mitigate the original strain.
\item The mutated strain $2$ is assumed to be $50\%$ more infectious, i.e.~$\calR_2=1.5\cdot \calR_1=1.275$ and hence spreads in time.
\item At the initial time ($25$ January) we assume that only $3\%$ of cases belong to strain $2$, i.e.~$z_0=0.03$.
\item Lab experiments~\cite{RKI_VOC210217} report in week $4$ around $5.6\%$ and in week $6$ already around $22\%$ of infections with strain $2$.
\end{enumerate}

Using the approximations~\eqref{E:T_star_lin} and~\eqref{E:Tz_min_lin} from the linearized model, strain $2$ will dominate strain $1$ at $T^\ast = - \frac{\ln z_0}{\calR_2-\calR_1} = 8.2 \equiv 41\,\mathrm{days}$. The minimal number of infected is expected to be $(x+y)_{\min} = 0.69\cdot x_0$ at time $\check{T} = 6.8 \equiv 34\,\mathrm{days}$.

\begin{figure}
\centerline{\includegraphics[width=.8\textwidth]{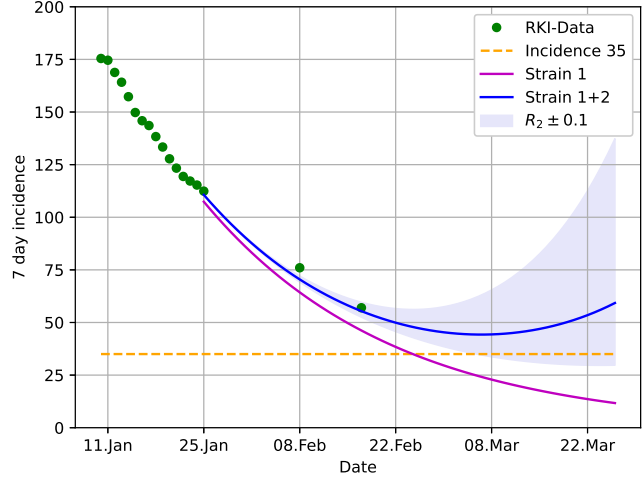}}
\caption{\label{F:Incidence} Incidence (per $100.000$ inhabitants in $7$ days) of strain $1$ (violet) and of both strains combined (blue) as predicted by the SIR--model~\eqref{E:SIR_nondim}. Parameters are $\calR_1=0.85$, $\calR_2=1.275$, $\gamma=1/5$. The green dots indicate incidences for entire Germany. The shaded area indicates the simulation range, if $\calR_2=1.275\pm 0.1$. The orange dash line indicates an incidence of $35$ that is viewed in Germany as a limit for relaxing the current lockdown.}
\end{figure}

\begin{figure}
\centerline{\includegraphics[width=.8\textwidth]{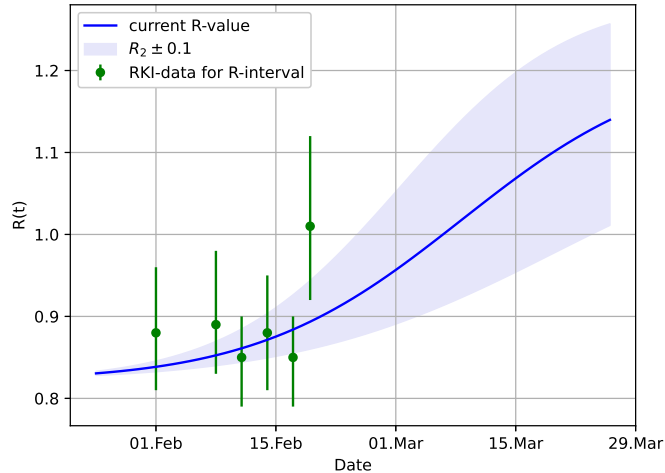}}
\caption{\label{F:currentR} Current value of the effective reproduction number~\eqref{E:R_of_T} for both strains. The green dots indicate the RKI--data for the $7$day--R0 together with the reported confidence interval. The shaded area indicates the simulation range, if $\calR_2=1.275\pm 0.1$.}
\end{figure}

The non--linear model~\eqref{E:SIR_nondim} cannot be solved analytically; hence we perform numerical simulations based on the parameter given above. Figures~\ref{F:Incidence},~\ref{F:currentR} show the dynamics of both strains and the current reproduction number based on Eqn.~\eqref{E:R_of_T}. The weekly incidences (new infections per $100.000$ inhabitants within $7$ days) are shown in Figure~\ref{F:Incidence}. The green dots indicate the reported data, see.~\cite{Datahub}. The violet curve shows the decay of the original strain $1$ with is reproduction number $\calR_1=0.85<1$. The blue curve shows the total incidence of both strains combined. At around 8 March, i.e.~$6$ weeks after the starting time of the simulation (25 January), the total number of infected reaches its minimum with an incidence of about $44$ per $100.000$. The time at which the minimum occurs is slightly larger than for the linear model ($41$ days for the nonlinear model compared to $34$ days for the linearized approximation). Our scenario and its parameters match quite well with the observed incidences, here shown for 8 and 17 February. The shaded area indicates the prediction uncertainty due to variations ($\calR_2\pm 0.1$) of the reproduction number of the second strain. Based on this simulation, the political target to push the infections below the threshold of $35$ before introducing relaxation measures seems questionable; at least on a short time horizon.

Figure~\ref{F:currentR} shows the current reproduction number for both strains combined as defined in Eqn.~\eqref{E:R_of_T}. Again, the green dots and error bars show the $7$--day reproduction number as reported by RKI in its daily situation reports~\cite{RKI_SitRep}. The blue curve shows our simulation results based on the non--linear model~\eqref{E:SIR_nondim} and again the shaded area indicates the uncertainty due to variation of the reproduction number of the second strain ($\calR_2\pm 0.1$). Currently, RKI is reporting an increase of the reproduction number breaking through the epidemic threshold of $\calR(t)=1$ as predicted by our model. The non--pharmaceutical interventions imposed by the government have not been altered in during the time span covered by the simulations, hence we may explain the increase of the overall reproduction number by growing influence of the second strain. 

\begin{figure}
\centerline{\includegraphics[width=.8\textwidth]{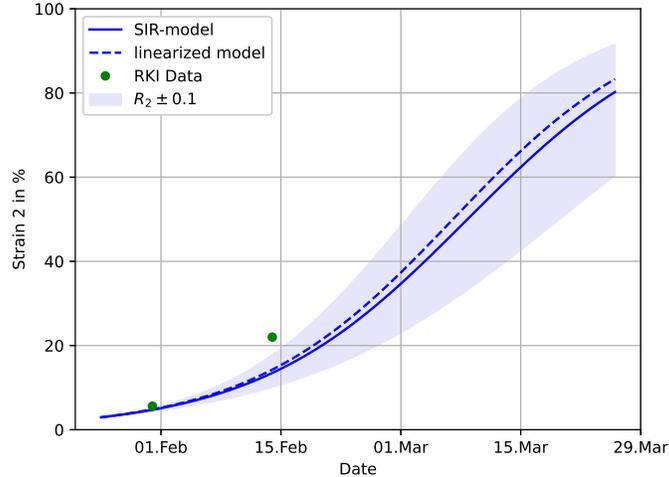}}
\caption{\label{F:vergleich_RKI}Simulation for of the relative share $p=y/(x+y)$ of the second strain. Reproduction numbers $\calR_1=0.85$, $\calR_2=1.275$. The green dots indicate the RKI data for the percentage of ''variants of concern'', see~\cite{RKI_VOC210217}. The solid blue line shows the non--linear SIR--Model~\eqref{E:SIR} whereas the dashed line is the linearized approximation. The shaded area indicates the simulation range of the non--linear SIR--model, if $\calR_2=1.275\pm 0.1$.}
\end{figure}

Figure~\ref{F:vergleich_RKI}  shows the relative share $p$ of strain 2 with respect to the total infections with SARS--COV--2 in Germany. The green dots indicate the data reported by RKI for week $4$ and week $6$, see~\cite{RKI_VOC210217}. The blue curves show our predictions; the dashed one corresponds to the approximation~\eqref{E:R_of_T_lin} in the linearized setting and the solid one corresponds to the non-linear SIR model. Both results do not differ significantly and the linearized model already predicts quite well the dynamics of the relative share of the second strain compared to all infections. Both results are within the range of the given data. Again, the shaded area indicates the uncertainty caused by variations in the reproduction number for the second strain. For the beginning of March we expect more than $40\%$ of infections with the second strain.

\section{Conclusions and Outlook}

In this work we have presented a two--strain SIR model to explain the spread of SARS--COV--2 variants like B.1.1.7 in Germany. For a linearized version of the model we were able to calculate relevant variables like the time of minimal infections or the dynamics of the share of variants analytically. These analytical approximations as well as simulations for the non--linear SIR model are compared to infection data reported by RKI. Our model shows a good level of agreement and gives rise to some concern regarding the near term future of the dynamics. For mid of March we expect to see in Germany a share of at least $40\%$ of variants. Moreover, the  figure of an incidence of $35$ per $100.000$ and $7$ days, which was introduced by politics as limit for easing the current lockdown measures, seems out of reach. 

In a follow--up study we will try to investigate the effect of the current ramping up of mass vaccinations. One might expect, that vaccinations will help to slow down the spread of the disease and hence force the level of incidence below a threshold that allows contact tracing by public health authorities.



\begin{thebibliography}{ab}
\bibitem{Pachetti} Pachetti, M., Marini, B., Benedetti, F., Giudici, F., Mauro, E., Storici, P., et.al. \emph{Emerging SARS--CoV--2 mutation hot spots include a novel RNA-dependent-RNA polymerase variant}. Journal of translational medicine, 18, 1-9. 2020.

\bibitem{Tang} Tang, J. W., Tambyah, P. A., \& Hui, D. S.  \emph{Emergence of a new SARS-CoV-2 variant in the UK}. The Journal of infection. 2020.

\bibitem{Korber} Korber, B., Fischer, W. M., Gnanakaran, S., Yoon, H., Theiler, J., Abfalterer, W., et.al.  \emph{Tracking changes in SARS--CoV--2 Spike: evidence that D614G increases infectivity of the COVID--19 virus}. Cell, 182(4), 812-827. 2020.


\bibitem{VOLZ} Volz, E., Mishra, S. and Chand, M., \emph{Transmission of SARS--CoV--2 lineage B. 1.1. 7 in England: insights from linking epidemiological and genetic data.} medRxiv [Preprint posted online January 4, 2021].


\bibitem{B117data}{Wikipedia: \emph{Development of the B.1.1.7.lineage} {\color{blue} \url{https://en.wikipedia.org/wiki/Lineage_B.1.1.7#Spread_in_Europe}}; access on 21.February.}

\bibitem{GHe20}{Götz, T.; Heidrich, P.: \emph{Early stage COVID-19 disease dynamics in Germany: models and parameter identification}. J. Math. in Industry. No. 10, (2020).}

\bibitem{HeS20}{Heidrich, P., Schäfer, M., Nikouei, M., Götz, T.: \textit{The COVID--19 outbreak in Germany --- Models and Parameter Estimation}. Commun. Biomath. Sci. 3, 37-59, 2020.}

\bibitem{SZ210205}{Sueddeutsche Zeitung: \emph{Die unsichtbare Welle}\newline {\color{blue} \url{www.sueddeutsche.de/wissen/coronavirus-mutante-b117-daten-1.5197700}}, 05.02.2021.}

\bibitem{RKI_VOC210217}{RKI: \emph{2. Bericht zu Virusvarianten von SARS--CoV--2 in Deutschland, insbesondere zur Variant of Concern (VOC) B.1.1.7} {\color{blue} \url{www.rki.de/DE/Content/InfAZ/N/Neuartiges_Coronavirus/DESH/Bericht_VOC_2021-02-17.pdf?__blob=publicationFile}}, 17.02.2020.}

\bibitem{Datahub}{COVID--19 Datenhub; time series of daily COVID--19 infections as reported by RKI, {\color{blue}\url{npgeo-corona-npgeo-de.hub.arcgis.com/datasets/6d78eb3b86ad4466a8e264aa2e32a2e4_0}}; access on 17.February.}

\bibitem{RKI_SitRep}{RKI: \emph{Daily Situation Reports}, {\color{blue} \url{www.rki.de/DE/Content/InfAZ/N/Neuartiges_Coronavirus/Situationsberichte/Gesamt.html}}.}



\end{thebibliography}
\end{document}